\documentstyle[preprint,eqsecnum,aps]{revtex}
\begin{document}
\draft
\preprint{McGill/94-43}
\title{A schematic model for fragmentation and phase transition \\
	in nuclear collisions}
\author{Jicai Pan \ and \ Subal Das Gupta }
\address{Department of Physics, McGill University \\
3600 University St., Montr\'{e}al, PQ, H3A 2T8 Canada}

\maketitle
\begin{abstract}
We develop here a simple yet versatile model for nuclear fragmentation
in heavy ion collisions. The model allows us to calculate
thermodynamic properties such as phase transitions as well
as the distribution of fragments at disassembly. In spite of its simplicity
the model gives very good fit to recent data taken at the Michigan National
Superconducting Cyclotron Laboratory. The model is an
extension of a lattice gas model which itself has strong overlaps
with percolation models which have been used in the past to
compare with nuclear fragmentation data.
\end{abstract}
\narrowtext

\newpage
\section{Introduction}

  Nuclear fragmentation has been intensively studied for the
purpose of measuring a possible liquid-gas phase transition
in nuclear matter \cite{curtin,bertsch}.
The most accessible quantity in experiments is a measurement of
the yield $Y(A)$ against $A$ where $A$ is the number of nucleons
in the composite that emerges from the collision \cite{finn,li1}.
A maximum in
the fluctuation in the fragment sizes could be an evidence for
critical phenomenon. In this paper we develop a schematic model for
nuclear fragmentation.  This is built on our previous work with a
lattice gas model \cite{pan} which in turn was closely linked with
much used percolation model for nuclear fragmentation
\cite{bauer,campi}.  The model we propose can be related to
phase transitions directly but we can also calculate the
yield $Y(A)$ against $A$. This yield against $A$ can be calculated
as a function of incident beam energy. There is just one parameter
in the model, namely,  the freeze-out density. This, however,
is approximately known from past experiences in
heavy ion collision theories. The model presented
here involves only moderate computing and can be enlarged to
include other effects.

\section{Thermodynamics of the model}

We consider a nearly central collision of approximately equal mass nuclei;
because of nucleon-nucleon collisions the system reaches thermal equilibrium.
We assume classical statistical mechanics is appropriate.  In that case the
canonical partition function of a $n$-particle system
can be written in separable form:
$Z({\rm can})=Z_p({\rm can})Z_r({\rm can})$
where $Z_p({\rm can})$ is given by
\begin{eqnarray}
Z_p({\rm can})\propto \int \exp [-\beta\sum_1^n{\bf p}_i^2/2m]
d^3{\bf p}_1.....d^3{\bf p}_n \ .
\end{eqnarray}
with $\beta$ being the inverse temperature and ${\bf p}_i$ being the
momentum of particle $i$.
The other part of the partition function is
\begin{eqnarray}
Z_{r}({\rm can})\propto \int \exp [-\beta \sum_{i<j}v({\bf r}_{ij})]
d^3{\bf r}_1......d^3{\bf r}_n . \label{zr}
\end{eqnarray}
Here the $v({\bf r}_{ij})$ is the potential between particles $i$ and $j$.
We approximate the configuration space part of the partition function
$Z_r({\rm can})$ by the partition function of the lattice gas model
\cite{pan,huang}.  The motivation for introducing the lattice gas model is that
it allows calculation of cluster distribution as in percolation model but
also allows calculation of thermodynamic properties including liquid-gas
phase transition in the same framework.

We consider a simple cube lattice. Each lattice site may have one particle
at the most.  For a system of $n$ particles the number of lattice sites,
$N$, should be greater than or equal to $n$.  When $N=n$ we have a
nucleus in its normal volume.  Thus the model considered here is limited to
normal volume and larger.  Cluster formation takes place at a freeze-out
density
which is expected to be about half of normal nuclear density so this is not a
debilitating limitation of the model.

We assume that the interaction is nearest neighbour type.  This reflects the
short range nature of nucleon-nucleon interaction.  The number of nearest
neighbours that a site has depends upon the dimension and the structure
of a lattice and will be denoted by $\gamma$.  For an infinite  simple
cube lattice we have $\gamma=6$. For finite systems that we
will consider here the $\gamma$ is less than 6 because of boundary effects.
The energy of nearest neighbour interaction
will be denoted by $-\epsilon $ where $\epsilon$ is positive and is related to
the binding energy.

We now obtain the equation of state for our system by first constructing the
partition function.  Let $N_{nn}$ be the number
of $nn$ bonds in a specific lattice configuration; the energy carried by these
bonds is then $-\epsilon N_{nn}$.  For a given $N$ and $n$
the partition function can be written as
\begin{eqnarray}
Z_r({\rm can})=\sum_{N_{nn}}g(N,n,N_{nn})e^{\beta\epsilon N_{nn}}
\end{eqnarray}
where $g(N,n,N_{nn})$ is a degeneracy factor satisfying
\begin{eqnarray}
\sum_{N_{nn}}g(N,n,N_{nn})=\frac{N!}{(N-n)!n!}
\end{eqnarray}
$Z_r({\rm can})$ was evaluated in \cite{pan} in the Bragg-Williams
approximation as well as in  Bethe-Peierls approximation which is
more accurate.  It will be  advantageous for us
to rederive the Bragg-Williams result.  In this approximation the number of
$nn$
bonds $N_{nn}$ is given and fixed when $N$ and $n$ are given.  If one site is
definitely occupied, the number of its $\gamma$ neighbours that are occupied
on the average is $\gamma n/N$.  Since there are $n$ nucleons in the system,
the number of $nn$ bonds is $\gamma n^2/2N$ where we have ensured that each
bond is counted only once and assumed that both $n$ and $N$ are so large that
boundary effects can be neglected. The partition function is then
\begin{eqnarray}
Z_r({\rm can})=\frac{N!}{(N-n)!n!}e^{\frac{1}{2}\beta \epsilon
 	\gamma \frac{n^2}{N}}
\end{eqnarray}
The equation of state can be calculated by using
$P=kT(\partial \ln Z({\rm can})/\partial V)_T
=kT(\partial \ln Z_r({\rm can})/\partial V)$ since
$Z_p({\rm can})$ has no $V$ dependence.  The volume $V$ is given by $V=a^3N$
with
$a^3=1/\rho_0=6.25$ fm$^3$ where $\rho_0$ is the normal nuclear density.  The
ground state volume of the nucleus consisting of $n$ nucleons is $V_0=a^3n$.
Using Stirling's formula for $N!$, $n!$ and $(N-n)!$ \ it is easy to show that
\begin{eqnarray}
P=\frac{kT}{a^3}\ln\frac{N}{N-n}-\frac{1}{2a^3}\epsilon \gamma (\frac{n}{N})^2
\end{eqnarray}
Using $n/N=V_0/V=\rho/\rho_0$ we finally get
\begin{eqnarray}
P=kT\rho_0\ln\frac{V}{V-V_0}-\frac{1}{2}
   \epsilon\rho_0\gamma(\frac{V_0}{V})^2 \label{state}
\end{eqnarray}
This equation of state has the same qualitative behaviour as the Van der Waals
gas.  For one mole of gas, the Van der Waals equation of state is
\begin{eqnarray}
P=\frac{N_AkT}{V-b}-\frac{a}{V^2}
\end{eqnarray}
The lattice gas pressure goes to infinity as $V$ approaches $V_0$.  The Van der
Waals gas pressure goes to infinity as $V$ is squeezed to the value $b$.
For large $V$ both the equations of state approach the perfect gas limit.
The critical point can be obtained analytically from equation (\ref{state}).
By setting
$\partial P/\partial \rho=\partial ^2P/\partial \rho^2=0$ at the critical
point we obtain $\rho_c=0.5 \rho_0$ and $kT_c=\gamma\epsilon/4$.  Better
calculations leave $\rho_c$ unchanged but reduces the value of $T_c$ to
$1.1275\epsilon$ \cite{stinch}.  An improved
calculation for the partition function was performed in \cite{pan} where
the equation of state of the lattice gas was compared with that for nuclear
matter in a mean field Skyrme interaction.  The qualitative behaviours are the
same.

\section{Cluster distribution}
As in percolation model calculations we simulate an event by Monte-Carlo
sampling and determine in each event the cluster distribution.  There
are two samplings to be done.  We have to determine which of the sites the
nucleons occupy and what their momenta are.  The two samplings
can be done independently of each other.  In filling up the $N$ sites with $n$
particles the Boltzmann factor (eq. \ref{zr}) needs to be taken into account.
We
do this in the following way.  Starting with a empty lattice we put the first
particle at random.  Once this has been put in, the $\gamma$ sites which are
immediate neighbours of this filled site are assigned a probability
$\propto \exp [\beta \epsilon ]$ whereas all other empty sites have probability
proportional to unity.  We now put the second particle according to this
probability distribution.  If at an intermediate step there are $m$ empty sites
we assign to each of these $m$ sites a probability proportional to
$\exp [q\beta \epsilon ]$ where $q$ is the number of its nearest neighbours
that
are already occupied.  The next filling is then done according to this
probability distributiion.  After the nucleons are all assigned their places
among the $N$ sites, the momentum of each particle is assigned
according to Maxwell-Boltzmann distribution.

  Now let us look at the distribution of clusters. In
a static lattice gas model in which the particles have no momenta,
because of binding energy two neighbouring nucleons are always bonded and
belong to the same cluster. The cluster is solely determined by this
spatial configuration, as in the case of a percolation model.

 In our model, however, two neighbouring particles may not form a
cluster because their relative kinetic energy might be higher than
the attracting potential and the bond might be broken.
It is natural to insist that two neighbouring nucleons are bonded if
the following condition is satisfied.
\begin{eqnarray}
{\bf p}_r^2/2\mu-\epsilon <0 \label{alg}
\end{eqnarray}
where ${\bf p}_r$ is the relative momentum and $\mu$ is the
reduced mass.

The frequency with which two nucleons appear in neighbouring sites
is mostly determined by the density.  The probability of ${\bf p}_r^2/2\mu$
exceeding the value $\epsilon$ increases with temperature since the momenta
of individual nucleons are obtained from Monte-Carlo sampling of
Maxwell-Boltzmann distribution at a given temperature.
Hence, the probability  that the two  neighbouring nucleons are
bonded decreases with increasing temperature, and the system becomes less
compact at higher temperature as it should.  A different parametrisation
used in \cite{li2} leads to similar effect.  It is also clear that one
important
feature of pure percolation models remains.  Provided the density is not too
small one will reach a percolation point in our schematic model at a certain
temperature.  This is
the point when for an infinite system one infinite cluster just appears.
For finite systems this signal for percolative phase transition
appears with a second
moment reaching a maximum which we discuss in the next section.

\section{Thermal and percolative phase transitions}

The power-law for emerging composites
$Y(A)\propto A^{-\tau}$
was first established by the Purdue group \cite{finn} in their
experimental data.  The Michigan experiment has identified \cite{li1}
the minimum in $\tau$ as a function of the bombarding energy in near
central collisions of two approximately equal ions.  At this minimum
the fluctuations are maximum indicating that this may be a critical point.
In the framework of the model we have introduced this will coincide with
a percolation point; does it also coincide with the
thermal critical point? The two types of critical points need not overlap.

  A percolation point can be found from the behaviour of the second moment
of cluster size defined as \cite{stauffer}:
\begin{eqnarray}
S_2={\sum_ss^2P(s)\over n}
\end{eqnarray}
where the $P(s)$ is the average number of clusters of size $s$. The
infinite cluster is excluded in the summation for an infinite system.
At the percolation point $S_2$ diverges in an infinite system
and is at maximum in a finite system. Campi \cite{campi} was the
first to exploit this feature by studying $S_2$ at different
numbers of fragments.

In our model (with cluster algorithm  given in \ref{alg})
the density determines the temperature at which the percolation
point is reached. In Fig. 1 we show  $S_2$ as a function of temperature
for $n$=85 (to simulate $^{40}$Ar+$^{45}$Sc collisions in \cite{li1})
on lattices $N$=$5^3$, $6^3$ and $7^3$ to simulate densities
$\rho=0.68\rho_0$, $0.39\rho_0$ and $0.25\rho_0$.
In the figure the largest cluster is excluded in every
event \cite{note1}.
If we assume that the density is close to the critical density
$\rho_c=0.5\rho_0$, then $S_2$ is maximum at the critical temperature $T_C$.
The minimum in $\tau$ is also observed at $T=T_C$ and $\rho=\rho_c$.
That is, the percolation point
coincides with the thermal critical point at the critical density.
Furthermore, it is found that
when the density increases, the temperature at which the percolation point
is reached increases and has the value of
$1.55T_C$ at $\rho=\rho_0$. This is in full agreement with the results
obtained from the Coniglio-Klein algorithm \cite{coniglio2} that was derived
mathematically by mapping the partition function into a percolation
model \cite{kastaleyn}. At this moment, it is not clear if there is any
{\it a priori } \ reason for this remarkable result to happen.

\section{Comparison with data}

We want to compare our model with the data presented in \cite{li1,li2}
where the data are fitted to a power-law and the deduced
exponent $\tau$ is plotted against the beam energy.  In our model
we can obtain the value of $\tau $ as a function of temperature. To
compare with experiments we need to relate the temperature to excitation
energy which then can be related to the beam energy.  This can be done
to different degrees of sophistication the simplest of which is to
assume completely classical limit.  In that case
there is no kinetic energy at zero temperature and
the ground state energy per nucleon is $-\epsilon N_{nn}^{max}/n$
where $N_{nn}^{max}$ is the maximal number of bonds possible.
For a system of 85 nucleons the
binding energy per nucleon is about 8.5 MeV. Since $N_{nn}^{max}$ is
determined by the geometry it can be used to fix the value
of $\epsilon$.  At the temperature $T$
the average energy per particle is
$1.5kT-\epsilon \overline{N}_{nn}/n$ where $\overline{N}_{nn}$,
the average value of $N_{nn}$, is obtained from the computer simulation.
We can then write
\begin{eqnarray}
\frac{3}{2}kT+\epsilon (N_{nn}^{max}-\overline{N}_{nn})/n
	=e^* \label{temp}
\end{eqnarray}
where $e^*$ is the excitation energy per nucleon. For equal mass
non-relativistic nuclear collisions we have
\begin{eqnarray}
e^*=E_{\rm beam}/4 \label{beam}
\end{eqnarray}
where $E_{\rm beam}$ is the beam energy per nucleon in the laboratory
(there is an implicit assumption here that all available
energy is thermalized).  Thus the temperature is related to the
beam energy.  Using this mapping we compare the
calculated $\tau$ (dashed line) with the experimental data in Fig. 2.
In the calculation we used $n=85$ and $N=6^3$ for freeze-out density
$\rho=0.39\rho_0$. The fit is
qualitatively correct but the $\tau$ increases too fast with beam energy.
This is clearly because we assumed full thermalization.
A much better quantitative fit is obtained if we
use the temperature extracted from experiments directly \cite{westfall,jacak}.
In this approach, the tail of the proton
spectrum in the laboratory is fitted by assuming that the proton has a
Maxwell-Boltzmann distribution in a source which is moving in the laboratory.
We take this mapping from the figure given in \cite{li2}.
The difference in mapping between temperature and beam energy in our
pure theoretical approach (c.f. eqs. \ref{temp} and \ref{beam})
and the phenomenological approach is shown in Fig. 3.

When this phenomenological mapping of temperature with beam energy is used
the fit between the experimental data taken from \cite{li1,li2,ogilvie}
and our  calculation is rather remarkable with one exception; the minimum
value of $\tau$
seen in experiment is lower than what we calculate.  It has been pointed out
that such low values of $\tau$ probably signify that fragmentation took
place in less compact geometry \cite{phair}.  For example if clusters
materialize from
a toroidal or a bubble-shaped configuration the calculated value of $\tau$
would be lower.  The determination of such shapes however require a
dynamical calculation. Quantum effect may also play an important role
at low temperatures.

  Fig. 4 repeats the calulation of Fig. 2 except that we assumed
at dissociation  $N=7^3$ leading to $\rho/\rho_0=0.25$.
The fit is not as good as that for $\rho/\rho_0=0.39$, especially the
minimum in $\tau$ is hardly discernible. Our
results indicate that $\rho/\rho_0=0.39$ is a better choice.

\section{Conclusions}
  The present model includes interactions between nucleons and
the momenta of nucleons. It is more realistic
than the percolation model which has been used in the past to look
for signatures of phase transition.  A significant feature
is that the same model encompasses thermal critical point and
percolation points. The very good agreement between our calculations
and the data taken at Michigan National Superconducting Cyclotron
is very encouraging for our model. We feel that existing data are
strongly suggestive of phase transition although more work needs
to be done to establish this more firmly. This includes
the measurements of various critical exponents and comparing them with
calculations using models such as ours. Work in this direction has
begun and is encouraging.

\acknowledgments

We wish to thank H. Guo, C. Gale, M. Grant and J. Kapusta for discussions
and W. Bauer for bringing Ref. \cite{li2} to our attention.
This research was supported in part by the Natural Sciences and Engineering
Council of Canada and in part by the FCAR fund of the Quebec Government.

\newpage
\section*{Figure captions}
\begin{description}
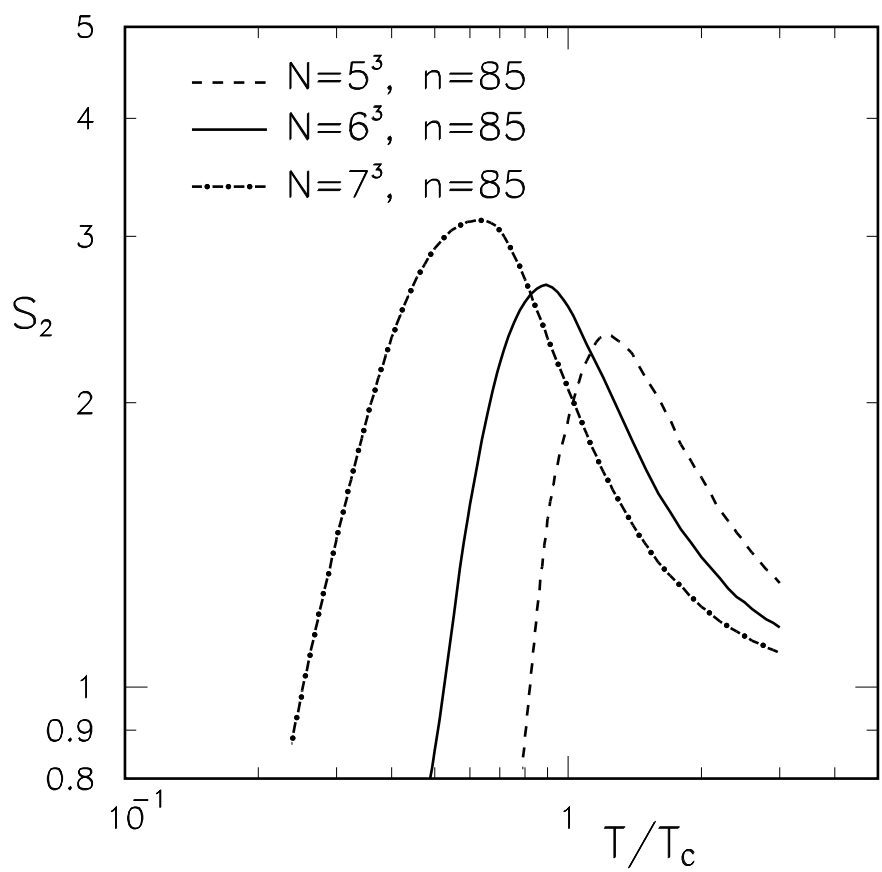
\item[Fig. 1]
The second moment $S_2$ is plotted as a function of $T/T_C$ at different
densities. Here $T_C=1.1275\epsilon$ is the critical temperature in a
lattice gas model.

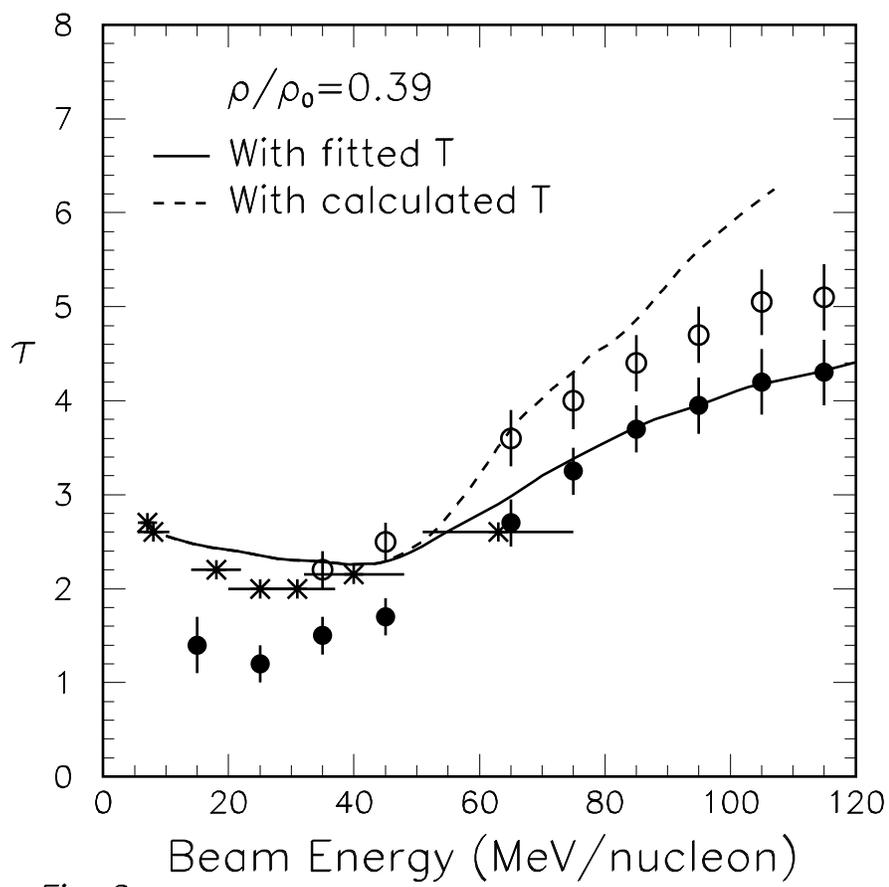
\item[Fig. 2]
The theoretical exponent $\tau$ is compared with experimental data
at different beam energies. The dashed curve is obtained by using the
temperature calculated from eqs. (\ref{temp}) and (\ref{beam}),
and the solid curve is
obtained by using the temperature fitted from experimental data.
The solid circles are the corrected data taken from \cite{li2},
the open circles are the uncorrected data taken from \cite{li1},
and the crosses are taken from \cite{ogilvie}.

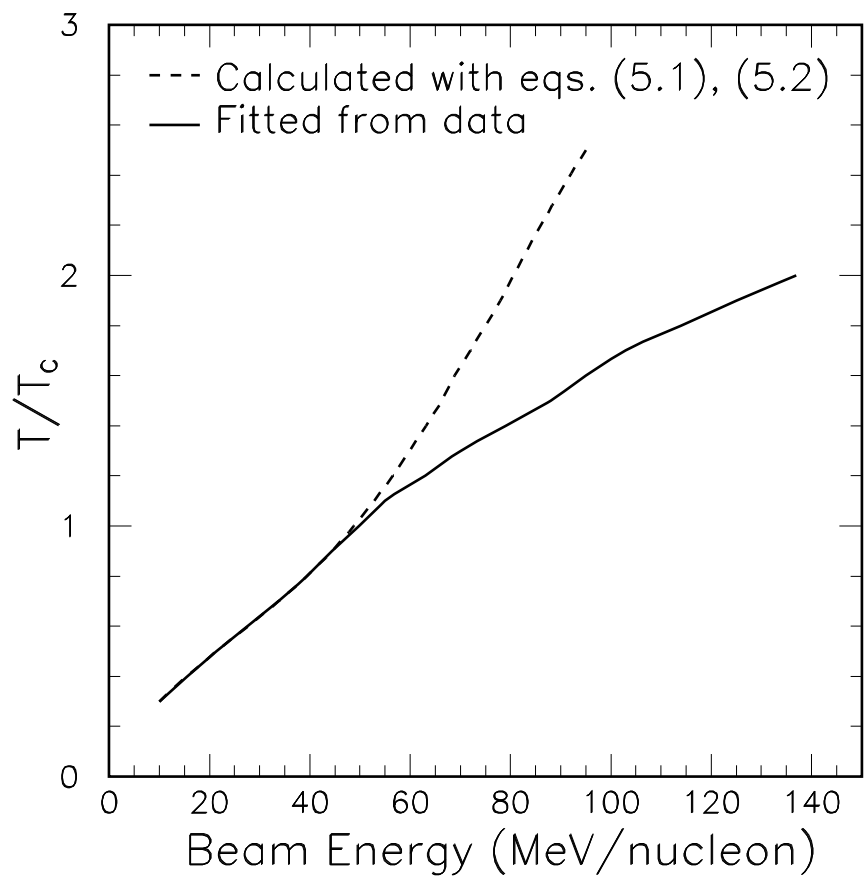
\item[Fig. 3]
The temperature is plotted as a function of beam energy.  The dashed curve
is calculated from eq. (\ref{temp}) and (\ref{beam}), and the solid curve
is taken from  \cite{li2} where the temperature is obtained by
fitting experimental data.

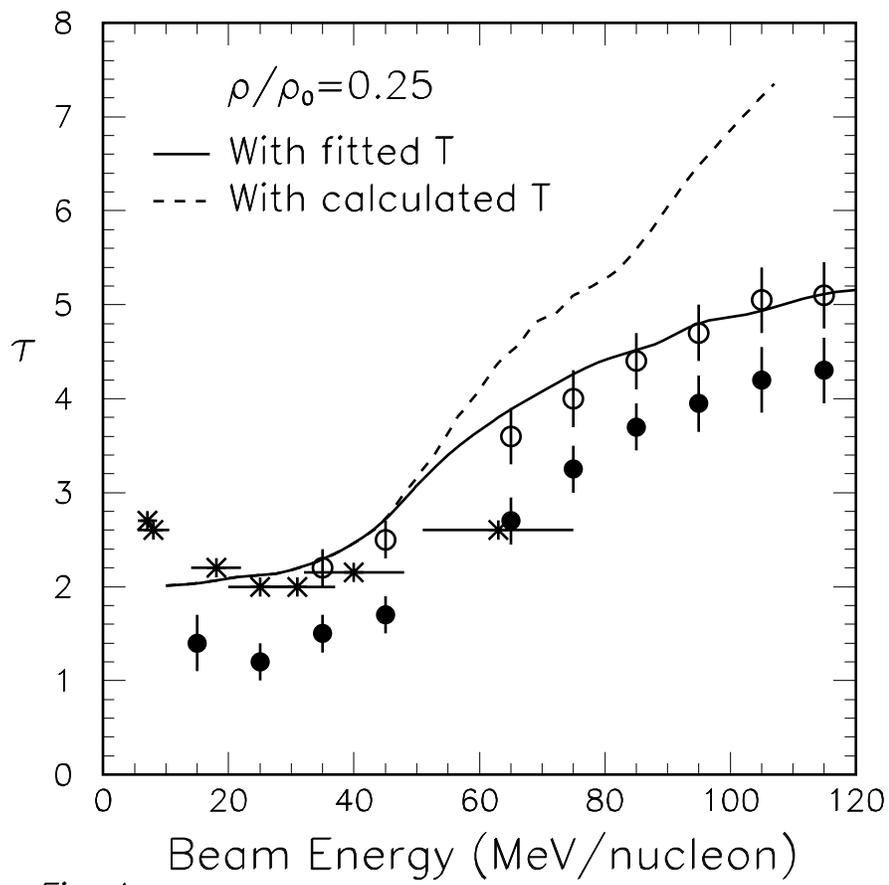
\item[Fig. 4]
The same as the Fig. 2, but for $\rho/\rho_0=0.25$.

\end{description}


\begin{references}
\bibitem{note1} In percolation theory the largest cluster is excluded in the
	percolated phase only \cite{stauffer}. In our simulations
	we studied $S_2$ by excluding the largest cluster in all events.
	We also studied $S_2$ by excluding the largest cluster only when it
	stretches from the bottom to the top, a method that better
	approximates a true percolation problem. We found
	little differences between the two methods for densities equal or
	higher than the critical density $\rho_c=0.5\rho_0$.
	When the density is much less than the critical density,
	excluding the largest cluster in all events no longer
	means leaving out a true percolation cluster.
\end{references}
\end{document}